\documentclass[%
print,
 amsmath,amssymb,
 aps,twocolumn
]{revtex4}

\usepackage{times}%此指令是令文章字体选用Palation.
\usepackage{graphicx}% Include figure files
\usepackage{subfigure}
\usepackage{dcolumn}% Align table columns on decimal point
\usepackage{bm}% bold math
\usepackage{dsfont}
\usepackage{mathrsfs}
\usepackage[landscape,papersize={297.1mm,210mm},left=1.9cm,right=1.6cm,top=2.7cm,bottom=2.8cm]{geometry}% 此指令是规定A4纸的页边距
\usepackage[                   %dvipdfm, pdflatex,pdftex这里决定运行文件的方式不同
            pdfstartview=FitH,
            colorlinks, %注释掉此项则交叉引用为彩色边框(将colorlinks和pdfborder同时注释掉)
            pdfborder=001,   %注释掉此项则交叉引用为彩色边框
            linkcolor=blue,
            anchorcolor=blue,
            citecolor=blue,
            urlcolor=blue
            ]{hyperref}
\usepackage{amsthm}
\usepackage{pifont}
\usepackage{diagbox}
\usepackage{multirow}
\makeatletter

\newcommand{\Rmnum}[1]{\expandafter\@slowromancap\romannumeral #1@}

\makeatother

\begin{document}
\title{Several families of entanglement criteria for multipartite quantum systems based on  the generalized Wigner-Yanase skew information and variance}

\author{Yan Hong$^{1,2}$}

\author{Xinlan Hao$^{1}$}

\author{Limin Gao$^{1,2}$}
\email{gaoliminabc@163.com}
\affiliation{$^1$ School of Mathematics and Science,  Hebei GEO University, Shijiazhuang 050031,  China \\
 $^2$   Intelligent Sensor Network Engineering Research Center of Hebei Province, Hebei GEO University, Shijiazhuang 050031,  China}

\begin{abstract}
Quantum entanglement plays a critical role in many quantum applications,  but detecting entanglement, especially in multipartite or high-dimensional quantum systems, remains a challenge. In this paper,  we propose several families of entanglement criteria for detecting  entanglement in multipartite or high-dimensional quantum states by the generalized Wigner-Yanase skew information $I^s(\rho,X)$ for $-1\leq s\leq0$ and variance.  We also reveal a complementary character between the criteria based on the generalized Wigner-Yanase skew information  and an alternative one based on variance through specific examples. We illustrate the merits of these criteria and show that the combination of the entanglement criteria has a stronger detection capability, as it is capable of detecting entangled states that remain unrecognized by other criteria.
\end{abstract}

\maketitle

\section{Introduction}
Quantum entanglement is a core concept in quantum theory and a key resource in quantum information processing and quantum computing \cite{EkertPRL1991,BennettWiesner1992,PRL70.1895,BennettDiVincenzo2000,GoYanLiEPL2008}.
The quantum entanglement structure of multipartite systems is more complex and can exhibit rich physical phenomena and application potential.
Therefore, the development of efficient detection  criteria for multipartite  entanglement is of great significance in gaining an in-depth understanding of the structure of multipartite quantum states and promoting the development of quantum technologies.

For bipartite quantum systems, numerous well-known criteria for identifying entanglement \cite{AggarwalAdhikariMajumdarPRA2024, MorelliHuberTavakoli2023,SchumacherAlberPRA2023,GhoshBosePRR2024,Bell1964,Peres1996,ChenWu2002,ChenWu2003,Rudolph2003,
GuhneHyllusGittsovichEisert2007,GittsovichGuhneHyllusEisert2008,LiLuo2013PRA,DohertyParrilo2002PRA,DohertyParrilo2004PRA}, such as Bell inequalities \cite{Bell1964}, the positive partial transposition criterion \cite{Peres1996}, the realignment criterion \cite{ChenWu2002,ChenWu2003,Rudolph2003}. Additionally, criteria based on different approaches, such as covariance matrix \cite{GuhneHyllusGittsovichEisert2007,GittsovichGuhneHyllusEisert2008},  quantum
Fisher information \cite{LiLuo2013PRA},  semidefinite program and the existence of extension \cite{DohertyParrilo2002PRA,DohertyParrilo2004PRA}, and more, have been developed to identify entanglement. However, for bipartite high-dimensional systems, these criteria are unable to identify  all entangled states. Therefore, further exploration is still required, even for entanglement detection in bipartite high-dimensional systems.

For multipartite quantum systems, determining whether a quantum state is entangled is more challenging due to the complexity of their structures. Over the past few years, a variety of multipartite entanglement detection methods using various approaches, have been proposed \cite{LiDuQinZhangDuZhouXiaoPRA2024,FrerotBaccariAcinPRX2022,PRA82.062113,
GaoYan2014,PLA401.127347,GuhneSeevinckNJP2019,XingHongGaoGaoYanQIP2024,DohertyParrilo2005PRA,SarbickiScalaChruscinski2020PRA,RenFan2023PRA,RicoHuber2024,
ZhangFei2019,AkbariAzhdargalam2019,Hyllus2012,Goth2012,HongLuoSong2015,AnnaBoschiPRB2023,LiWangFeiJost2014PRA,
HassanJoag2007QIC,TothVertesi2018,GessnerSmerziPRA2016}.
Doherty $et$ $al$. introduced  a detection method for multipartite entanglement that utilizes the existence of extending the state to a larger space consisting of some copies of each of the subsystems \cite{DohertyParrilo2005PRA}.
In Ref. \cite{SarbickiScalaChruscinski2020PRA},  a series of entanglement criteria based on correlation matrix have been proposed, including some previous results of  Refs. \cite{LiWangFeiJost2014PRA, HassanJoag2007QIC} as  special cases.
Based on the generalized state-dependent entropic uncertainty relations for multiple measurements,
the work of Ref. \cite{RenFan2023PRA} has presented a strategy for identifying entanglement in GHZ states ranging from three to ten qubits through the Quafu cloud quantum computing platform.
The paper Ref. \cite{RicoHuber2024} has established a systematic method for nonlinear entanglement detection by trace polynomial inequalities. Through this approach, bipartite witnesses can be used to detect multipartite entanglement. Furthermore, it emphasizes the advantage of the nonlinear detection method, which is capable of successfully identifying  pairs of entangled states and acting as a witness that cannot be judged by linear detection methods. Quantum Fisher information  is a pivotal concept in quantum metrology and plays a significant role in various fields such as parameter estimation, entanglement detection,  and quantum information geometry. Since separable states tend to decrease the upper bound of quantum Fisher information, this characteristic endows quantum Fisher information with important applications in entanglement detection. By measuring quantum Fisher information of quantum states, several methods for entanglement detection have been developed \cite{ZhangFei2019,AkbariAzhdargalam2019,Hyllus2012,Goth2012,HongLuoSong2015,AnnaBoschiPRB2023,TothVertesi2018,GessnerSmerziPRA2016}.

In this paper, we mainly investigate the detection of multipartite entanglement via the  generalized Wigner-Yanase skew information and variance. The structure of the paper is as follows. In Sec. II, we  review  the notions and  some basic features of the generalized Wigner-Yanase skew information $I^s(\rho,X)$ and variance. In Sec. III, we present several families of entanglement detection  criteria from the perspective of the generalized Wigner-Yanase skew information  for $-1\leq s\leq0$ and variance. These criteria are classified into two categories, which are based on mutually unbiased measurements and general symmetric informationally complete measurements, respectively. We also illustrate the complementary features of these criteria based on the generalized Wigner-Yanase skew information for $-1\leq s\leq0$ and variance. In Sec. IV, the paper is concluded and discussed.

\section{ the generalized Wigner-Yanase skew information vs variance }

The concept of  generalized Wigner-Yanase skew information is closely related with the function $f_s(a,b)$
 introduced in  Ref. \cite{Hardy1952} as follows,
\begin{equation*}
 f_s(a,b)=\left\{\begin{array}{ll} \Big(\dfrac{a^s+b^s}{2}\Big)^{1/s}, & \quad \textrm{ if } a>0,b>0,\\
0, & \quad \textrm{ if } a=0 \textrm{ or }  b=0,
\end{array}
\right.
\end{equation*}
when $s\in(-\infty,0)$;
$f_0(a,b)=\lim\limits_{s\rightarrow0}f_s(a,b)=\sqrt{ab}$ when $s=0$;
$f_{-\infty}(a,b)=\lim\limits_{s\rightarrow-\infty}f_s(a,b)=\min\{a,b\}$ when $s=-\infty$.
For the above function $ f_s(a,b)$,  it increases monotonically with $s$ \cite{Hardy1952}.

For quantum state $\rho$ with the spectral decomposition
$\rho=\sum\limits_l\lambda_l|\psi_l\rangle\langle\psi_l|$, the
 generalized Wigner-Yanase skew information of the observable $X$ can be expressed as \cite{YangQiao2022}
\begin{equation*}
\begin{aligned}
I^s(\rho,X):=&\textrm{Tr}(\rho X^2)-\sum\limits_{l, l'}f_s(\lambda_l,\lambda_{l'})|\langle\psi_l|X|\psi_{l'}\rangle|^2\\
=&\sum\limits_{l\neq l'}[\lambda_l-f_s(\lambda_l,\lambda_{l'})]|\langle\psi_l|X|\psi_{l'}\rangle|^2.
\end{aligned}
\end{equation*}
When $-1\leq s\leq0$, the generalized Wigner-Yanase skew information $I_s(\rho,X)$ reduces to the metric adjusted skew information \cite{YangQiao2022}.
Specifically, when the value of $s$ is $-1$ or $0$,  $I^{-1}(\rho,X)$ and $I^0(\rho,X)$ represent two different information metrics, namely quantum Fisher information and Wigner-Yanase skew information, respectively.

The variance of the observable $X$ in the state $\rho$ is defined as $V(\rho,X)=\textrm{Tr}(\rho X^2)-\big[\textrm{Tr}(\rho X)\big]^2$ \cite{HofmannTakeuchi2003}.
It should be pointed out that $I^s(\rho,X) =V(\rho,X)$ for any pure
state $\rho$, but  $I^s(\rho,X) \leq V(\rho,X)$ for the mixed state $\rho$.

The generalized Wigner-Yanase skew information and variance have some important properties, which will be useful in the subsequent discussion \cite{YangQiao2022,HofmannTakeuchi2003,GuhnePRL2004,FrankHansen2008}.

(1) (Monotonicity of $s$) For any observable $X$,  the generalized Wigner-Yanase skew information is  monotonically decreasing with respect to $s$, that is,
$$ I^0(\rho,X)\leq\cdots\leq I^{-\infty}(\rho,X)\leq V(\rho,X),$$
where the  equality holds for any pure states.

(2) (Convexity or Concavity) For any observable $X$, the
 generalized Wigner-Yanase skew information is convex  when $-1\leq s\leq0$ and the variance is concave, that is,
\begin{equation*}
\begin{aligned}
I^s(\sum\limits_ip_i\rho_i,X)&\leq\sum\limits_ip_iI^s(\rho_i,X), \\
V(\sum\limits_ip_i\rho_i,X)&\geq\sum\limits_ip_iV(\rho_i,X),
\end{aligned}
\end{equation*}
when $-1\leq s\leq0$.  Here $p_i\geq 0$ and $\sum\limits_ip_i=1$.

(3) (Additivity) For any $N$-partite quantum  state $\bigotimes\limits_{i=1}^N|\psi_i\rangle$ in an $N$-partite quantum system $\mathcal{H}_1\otimes \mathcal{H}_2\otimes\cdots \otimes \mathcal{H}_N$,  the generalized Wigner-Yanase skew information and  the variance hold
\begin{equation*}
\begin{aligned}
&I^s\Big (\bigotimes\limits_{i=1}^N|\psi_i\rangle,\sum\limits_{i=1}^N\mathbb{X}_i\Big )
=V\Big (\bigotimes\limits_{i=1}^N|\psi_i\rangle,\sum\limits_{i=1}^N\mathbb{X}_i\Big )\\
&=\sum\limits_{i=1}^NV(|\psi_i\rangle,X_i),
\end{aligned}
\end{equation*}
where $|\psi_i\rangle$ is the quantum state on subsystem $\mathcal{H}_i$ and the operator  $\sum\limits_{i=1}^N\mathbb{X}_i$ is defined as
\begin{equation}\label{symbol1}
\begin{aligned}
\sum\limits_{i=1}^N\mathbb{X}_i=\sum\limits_{i=1}^N\textbf{1}_1\otimes\cdots\otimes \textbf{1}_{i-1}\otimes X_i\otimes \textbf{1}_{i+1}\otimes\cdots\otimes \textbf{1}_N
\end{aligned}
\end{equation}
 with $X_i$ being an observable acting on the subsystem $\mathcal{H}_i$ and $\textbf{1}_j$ being the identity matrix acting on the subsystem $\mathcal{H}_j$.

\section{ Entanglement criteria from the generalized Wigner-Yanase skew information for $-1\leq s\leq0$ and variance}

In an $N$-partite quantum system $\mathcal{H}_1\otimes \mathcal{H}_2\otimes\cdots \otimes \mathcal{H}_N$,  a pure state $|\Psi\rangle$ is fully separable if it can be written as a product state of all parties, that is, $|\Psi\rangle=\bigotimes\limits_{i=1}^N|\psi_i\rangle$ with $|\psi_i\rangle$ being the substate of subsystem $\mathcal{H}_i$  \cite{HorodeckiRMP2009,GuhneToth2009}.
An $N$-partite mixed state  $\rho$ is said to be fully separable if it can be expressed as a convex combination of
fully separable pure state: $\rho=\sum\limits_lp_l|\Psi^{(l)}\rangle\langle\Psi^{(l)}|$ where the pure state $|\Psi^{(l)}\rangle$ is fully separable  \cite{HorodeckiRMP2009,GuhneToth2009}. Otherwise, the quantum state is called entangled.

In order to  review our  entanglement criteria based on the generalized Wigner-Yanase skew information for $-1\leq s\leq0$ and variance, we categorize the observables in the entanglement criteria into the following two categories: mutually unbiased measurements and general symmetric informationally complete measurements.

\subsection{Entanglement criteria involving  mutually unbiased measurements }

These measurements on Hilbert space $\mathcal{H}$ with $\textrm{dim}(\mathcal{H})=d$,
$\mathcal{M}^{(u)}=\{M^{(uv)}|M^{(uv)}\geq0,\sum\limits_{v=1}^{d}M^{(uv)}=\textbf{1} \}$, ($1\leq u\leq m$), with $d$ elements each,  are  mutually unbiased measurements (MUMs)  if and only if \cite{KalevGour2014,LaiLuoPRA2022}
\begin{equation*}
\begin{aligned}
\textrm{Tr}(M^{(uv)})=&1,\\
\textrm{Tr}(M^{(uv)}M^{(u'v')})=&\delta_{uu'}\delta_{vv'}\kappa+(1-\delta_{vv'})\delta_{uu'}\dfrac{1-\kappa}{d-1}\\
&+(1-\delta_{uu'})\dfrac{1}{d},
\end{aligned}
\end{equation*}
where $\dfrac{1}{d}< \kappa \leq1$, and  $\kappa=1$ if and only if all $M^{(uv)}$ are rank 1 projectors given by mutually unbiased bases.
The study of Ref. \cite{KalevGour2014} has not only provided a method for constructing a complete set of MUMs ( that is, $m=d+1$ MUMs), but also pointed out that any complete set of MUMs can be obtained in this way.

\emph{Theorem 1}. For an $N$-partite quantum system $\mathcal{H}_1\otimes \mathcal{H}_2\otimes\cdots \otimes \mathcal{H}_N$ with $\textrm{dim}(\mathcal{H}_i) = d$,
let  $\mathcal{M}^{(u)}=\{M^{(uv)}|M^{(uv)}\geq0,\sum\limits_{v=1}^{d}M^{(uv)}=\textbf{1} \}$  ($1\leq u\leq d+1$) be  a complete set of MUMs of  Hilbert space $\mathcal{H}_i$
and $\sum\limits_{i=1}^N\mathbb{M}_i^{(uv)}=\sum\limits_{i=1}^N\textbf{1}_1\otimes\cdots\otimes \textbf{1}_{i-1}\otimes M_i^{(uv)}\otimes \textbf{1}_{i+1}\otimes\cdots\otimes \textbf{1}_N$ with $M^{(uv)}_i=M^{(uv)}$ for any $i=1,2,\cdots,N$.
 Based on the above symbols and concepts,
for any fully separable state $\rho$, it holds that

(I) (the generalized Wigner-Yanase skew information criterion)
\begin{equation}\label{MUMs-SI}
\begin{aligned}
\sum\limits_{u=1}^{d+1}\sum\limits_{v=1}^dI^s\Big(\rho,\sum\limits_{i=1}^N\mathbb{M}_i^{(uv)}\Big)
\leq N\kappa d-N;
\end{aligned}
\end{equation}

(II) (variance criterion)
\begin{equation}\label{MUMs-SII}
\begin{aligned}
\sum\limits_{u=1}^{d+1}\sum\limits_{v=1}^dV\Big(\rho,\sum\limits_{i=1}^N\mathbb{M}_i^{(uv)}\Big)
\geq N\kappa d-N.
\end{aligned}
\end{equation}
Here $-1\leq s\leq0$.
Consequently, if a quantum state $\rho$ violates the above inequalities, it is entangled.
In addition, both inequality (\ref{MUMs-SI}) and inequality (\ref{MUMs-SII}) hold with equality for fully separable pure states.
The proof of Theorem 1 is given in Appendix A.

Considering the relation between the generalized Wigner-Yanase skew information and variance, we show that the entanglement
criteria based on the generalized Wigner-Yanase skew information for $-1\leq s\leq0$ and variance are complementary to each other.
Noting that the bounds of the generalized Wigner-Yanase skew information and variance are the same in Theorem 1,  we conclude as follows:

(A1) If $\sum\limits_{u=1}^{d+1}\sum\limits_{v=1}^dI^s\Big(\rho,\sum\limits_{i=1}^N\mathbb{M}_i^{(uv)}\Big)\leq
\sum\limits_{u=1}^{d+1}\sum\limits_{v=1}^dV\Big(\rho,\sum\limits_{i=1}^N\mathbb{M}_i^{(uv)}\Big)
< N\kappa d-N$, then quantum state $\rho$ is entangled and can be  identified by entanglement
criteria of Theorem 1 based on variance.

(A2) If $N\kappa d-N<\sum\limits_{u=1}^{d+1}\sum\limits_{v=1}^dI^s\Big(\rho,\sum\limits_{i=1}^N\mathbb{M}_i^{(uv)}\Big)\leq
\sum\limits_{u=1}^{d+1}\sum\limits_{v=1}^dV\Big(\rho,\sum\limits_{i=1}^N\mathbb{M}_i^{(uv)}\Big)$, then quantum state $\rho$ is entangled and can be  identified by entanglement
criteria of Theorem 1 based on the generalized Wigner-Yanase skew information for $-1\leq s\leq0$.

(A3) If $\sum\limits_{u=1}^{d+1}\sum\limits_{v=1}^dI^s\Big(\rho,\sum\limits_{i=1}^N\mathbb{M}_i^{(uv)}\Big)
\leq N\kappa d-N\leq
\sum\limits_{u=1}^{d+1}\sum\limits_{v=1}^dV\Big(\rho,\sum\limits_{i=1}^N\mathbb{M}_i^{(uv)}\Big)$,
 then neither the
entanglement criteria of Theorem 1 based on  the generalized Wigner-Yanase skew information for $-1\leq s\leq0$
 nor that based on variance  can  ascertain whether or not $\rho$ is entangled.

From (A1) and (A2), we find the complementarity of inequalities (\ref{MUMs-SI}) and (\ref{MUMs-SII}). That is, inequality (\ref{MUMs-SI}) can identify some entangled states undetected by inequality (\ref{MUMs-SII}), and vice versa.
Next, we will elaborate on the complementary relationship between inequalities (\ref{MUMs-SI}) and (\ref{MUMs-SII}) through specific examples. Furthermore, we will highlight that the combination is capable of detecting entangled states that remain unrecognized by other criteria.

$\emph{Example 1}.$  Consider the family of $N$-qubit quantum states mixed by Dicke state and white noise,
 $$\rho_1(p)=p|D_N\rangle\langle D_N|+\dfrac{1-p}{2^N}\textbf{1},$$
 where $$|D_N\rangle=\sqrt{\dfrac{(\frac{N}{2})!(\frac{N}{2})!}{N!}}\sum\limits_{i_1+\cdots+i_N=\frac{N}{2}}|i_1i_2\cdots i_N\rangle$$ when $N$ is even,
$$|D_N\rangle=\sqrt{\dfrac{(\frac{N-1}{2})!(\frac{N+1}{2})!}{N!}}\sum\limits_{i_1+\cdots+i_N=\frac{N+1}{2}}|i_1i_2\cdots i_N\rangle$$ when $N$ is odd with $i_1,\cdots, i_N=0$ or 1.

For our Theorem 1, let $s=-1$, $\kappa=1$ and MUMs are given in Appendix C for $d=2$.
Using inequality (\ref{MUMs-SI}) for $N=3,4,5,8,9$,  we find that $\rho_1(p)$ is entangled state for $0.5254<p\leq1,$ $0.3333<p\leq1,$ $0.3312<p\leq1,$
$0.1269<p\leq1,$ $0.1868<p\leq1,$ respectively. But inequality (\ref{MUMs-SII})  can not detect any entangled state.
So, inequality (\ref{MUMs-SI}) is more powerful than both  inequality (\ref{MUMs-SII}).
At the same time,   for $N=3,4,5,8,9$, we can find that the parameter range of entangled states determined by Proposition 1 of Ref. \cite{HongLuoSong2015} is $0.5254<p\leq1,$ $0.3967<p\leq1,$ $0.3312<p\leq1,$
$0.2060<p\leq1,$ $0.1868<p\leq1,$ respectively. Hence, inequality (\ref{MUMs-SI}) is more powerful than
Proposition 1 of Ref. \cite{HongLuoSong2015} in detecting entangled states  for $N=4,8$.

$\emph{Example 2}.$ Consider a family of $N$-partite quantum states in $\mathcal{H}_1\otimes \mathcal{H}_2\otimes\cdots \otimes \mathcal{H}_N$ with $\textrm{dim}(\mathcal{H}_i) = N$, that is,
$$\rho_2(p)=p|S_N\rangle\langle S_N|+\dfrac{1-p}{N^N}\textbf{1},$$
which is mixed by quantum state $|S_N\rangle$ and white noise. Here $$|S_N\rangle=\dfrac{1}{\sqrt{N!}}\sum\limits_\sigma(-1)^{\textrm{sgn}(\sigma)}|\sigma\rangle,$$ where $\textrm{sgn}(\sigma)$ is the signature of the permutation of $01\cdots (N-1)$,
and the sum is taken over all permutations.

For our Theorem 1, let $s=-1$, $\kappa=1$ and MUMs are given in Appendix C for $d=3,4,5,8,9$.
We use inequality (\ref{MUMs-SII}) to detect  entanglement of $\rho_2(p)$  for $N=3,4,5,8,9$, and we conclude that  the parameter range
of entangled states  detected by inequality (\ref{MUMs-SII}) is $\dfrac{1}{4}<p\leq1,$ $\dfrac{1}{5}<p\leq1,$ $\dfrac{1}{6}<p\leq1,$
$\dfrac{1}{9}<p\leq1,$ $\dfrac{1}{10}<p\leq1,$ respectively.
However, neither inequality (\ref{MUMs-SI}) nor Proposition 1 of Ref. \cite{HongLuoSong2015} can detect any entangled state.
Hence, we can see that inequality (\ref{MUMs-SII}) has a stronger ability to detect entanglement than both  inequality (\ref{MUMs-SI}) and Proposition 1 of Ref. \cite{HongLuoSong2015}.

From Examples 1 and 2, we can find that the detection capabilities of entanglement criteria based on inequality (\ref{MUMs-SI}) and inequality (\ref{MUMs-SII}) vary for different quantum states, and  they are complementary to each other, as summarized in Table I. The combination of
the two entanglement criteria has a stronger detection ability, that is, it  can detected some entangled states that cannot be detected by Proposition 1 of Ref. \cite{HongLuoSong2015}, which is also summarized in Table I.
\begin{table*}
\caption{\label{tab:table1} {In this table,  the symbol \textquotedblleft\ding{53}\textquotedblright~indicates that the corresponding criterion is unable to detect any entangled state.
For the quantum
states $\rho_1(p)$, inequality (\ref{MUMs-SI}) based on the generalized Wigner-Yanase skew information for $-1\leq s\leq0$ can always
 detect some entangled states, but inequality (\ref{MUMs-SII}) based on variance cannot. Moreover, for $\rho_1(p)$, inequality (\ref{MUMs-SI})  can  always detect more entangled states than Proposition 1 of Ref. \cite{HongLuoSong2015} for $N=4,8$; inequality (\ref{MUMs-SI}) and Proposition 1 of Ref. \cite{HongLuoSong2015}  have the same entanglement detection capability for $N=3,5,9$.
 For the quantum
states $\rho_2(p)$, inequality (\ref{MUMs-SII}) can always detect some
entangled states, but neither
inequality (\ref{MUMs-SI})  nor  Proposition 1 of Ref. \cite{HongLuoSong2015} cannot.}}

\vskip 0.2cm
\begin{tabular}{*{11}{c}}
\hline\hline
\multirow{2}{*}{{\shortstack{States}}}&\multicolumn{2}{c}{$N=3$}&\multicolumn{2}{c}{$N=4$}&\multicolumn{2}{c}{$N=5$}&\multicolumn{2}{c}{$N=8$}&\multicolumn{2}{c}{$N=9$}\\
\cline{2-11}
&$\rho_1(p)$&$\rho_2(p)$&$\rho_1(p)$&$\rho_2(p)$&$\rho_1(p)$&$\rho_2(p)$&$\rho_1(p)$&$\rho_2(p)$&$\rho_1(p)$&$\rho_2(p)$\\
\cline{1-11}\\
\multicolumn{1}{l}{\textrm{Inequality} (\ref{MUMs-SI}) }&\multicolumn{1}{l}{$p>0.5254$}&\multicolumn{1}{l}{\quad\ding{53} }&\multicolumn{1}{l}{$p>0.3333$}&\multicolumn{1}{l}{\quad\ding{53} }&\multicolumn{1}{l}{$p>0.3312$}&\multicolumn{1}{l}{\quad\ding{53} }&\multicolumn{1}{l}{$p>0.1269$ }&\multicolumn{1}{l}{\quad\ding{53} }&\multicolumn{1}{l}{$p>0.1868$}&\multicolumn{1}{l}{\quad\ding{53} }\\\\
\multicolumn{1}{l}{\textrm{Inequality} (\ref{MUMs-SII}) }&\multicolumn{1}{l}{\quad\quad\ding{53} }&\multicolumn{1}{l}{$p>\frac{1}{4}$ }&\multicolumn{1}{l}{\quad\quad\ding{53} }&\multicolumn{1}{l}{$p>\frac{1}{5}$ }&\multicolumn{1}{l}{\quad\quad\ding{53} }&\multicolumn{1}{l}{$p>\frac{1}{6}$ }&\multicolumn{1}{l}{\quad\quad\ding{53} }&\multicolumn{1}{l}{ $p>\frac{1}{9}$ }&\multicolumn{1}{l}{\quad\quad\ding{53} }&\multicolumn{1}{l}{$p>\frac{1}{10}$ }\\\\
\multicolumn{1}{l}{\textrm{Proposition 1 of Ref.}\cite{HongLuoSong2015}}  &\multicolumn{1}{l}{$p>0.5254$ }&\multicolumn{1}{l}{\quad\ding{53} }&\multicolumn{1}{l}{$p>0.3967$ }&\multicolumn{1}{l}{\quad\ding{53} }&\multicolumn{1}{l}{$p>0.3312$ }&\multicolumn{1}{l}{\quad\ding{53} }&\multicolumn{1}{l}{ $p>0.2060$ }&\multicolumn{1}{l}{\quad\ding{53} }&\multicolumn{1}{l}{$p>0.1868$ }&\multicolumn{1}{l}{\quad\ding{53}  }\\
\hline\hline
\end{tabular}
\end{table*}

\subsection{Entanglement criteria involving general symmetric informationally complete measures}

In Hilbert space $\mathcal{H}$ with $\textrm{dim}(\mathcal{H})=d$, if a positive operator-valued measures (POVM) $\{G^{(u)}:u=1,\cdots,d^2\}$ satisfies $\textrm{Tr}\big[(G^{(u)})^2\big]=\eta$ and $\textrm{Tr}(G^{(u)}G^{(u')})=\dfrac{1-d\eta}{d(d^2-1)}$ for any $u\neq u'$, then it is called general symmetric informationally complete measures (GSICs), where $\dfrac{1}{d^3}<\eta\leq\dfrac{1}{d^2}$ \cite{GourKalev2014,LaiLuoPRA2022} and $\eta=\dfrac{1}{d^2}$ if and only if all $G^{(u)}$ are rank 1 projectors given by symmetric informationally complete
positive operator-valued measures. The paper Ref. \cite{GourKalev2014} has provided a method of constructing a GSICs and indicated that any GSICs can be constructed by this method.

\emph{Theorem 2}.
For an $N$-partite quantum system $\mathcal{H}_1\otimes \mathcal{H}_2\otimes\cdots \otimes \mathcal{H}_N$ with $\textrm{dim}(\mathcal{H}_i) = d$,
let  $\{G^{(u)}:u=1,\cdots,d^2\}$ be  GSIC of  Hilbert space $\mathcal{H}_i$ and
$\sum\limits_{i=1}^N\mathbb{G}_i^{(u)}=\sum\limits_{i=1}^N\textbf{1}_1\otimes\cdots\otimes \textbf{1}_{i-1}\otimes G_i^{(u)}\otimes \textbf{1}_{i+1}\otimes\cdots\otimes \textbf{1}_N$ with $G^{(u)}_i=G^{(u)}$ for any $i=1,2,\cdots,N$.
 Based on the above symbols and concepts,  we can obtain the criteria for identifying entanglement via the generalized Wigner-Yanase skew information  for $-1\leq s\leq0$ and variance.
For any fully separable state $\rho$, it holds that

(i) (the generalized Wigner-Yanase skew information criterion)
\begin{equation}\label{GSIC-SI}
\begin{aligned}
\sum\limits_{u=1}^{d^2}I^s\Big(\rho,\sum\limits_{i=1}^N\mathbb{G}_i^{(u)}\Big)
\leq  Nd\eta-N\dfrac{\eta d^2+1}{d(d+1)};
\end{aligned}
\end{equation}

(ii) (variance criterion)
\begin{equation}\label{GSIC-SII}
\begin{aligned}
\sum\limits_{u=1}^{d^2}V\Big(\rho,\sum\limits_{i=1}^N\mathbb{G}_i^{(u)}\Big)
\geq Nd\eta-N\dfrac{\eta d^2+1}{d(d+1)}.
\end{aligned}
\end{equation}
Here $-1\leq s\leq0$. Accordingly, any violation of the above inequalities indicates that $\rho$ is entangled.
Moreover, we should point out that both inequality (\ref{GSIC-SI}) and inequality (\ref{GSIC-SII}) hold with equality for fully separable pure states.
The proof of Theorem 2 is given in Appendix B.

Next, we show that the entanglement
criteria based on the generalized Wigner-Yanase skew information for $-1\leq s\leq0$ and variance in Theorem 2,  are complementary to each other, and we conclude  as follows:

(B1) If $\sum\limits_{u=1}^{d^2}I^s\Big(\rho,\sum\limits_{i=1}^N\mathbb{G}_i^{(u)}\Big)\leq
\sum\limits_{u=1}^{d^2}V\Big(\rho,\sum\limits_{i=1}^N\mathbb{G}_i^{(u)}\Big)
< Nd\eta-N\dfrac{\eta d^2+1}{d(d+1)}$, then quantum state $\rho$ is entangled and can be  detected by entanglement
criteria of Theorem 2 based on variance.

(B2) If $Nd\eta-N\dfrac{\eta d^2+1}{d(d+1)}<\sum\limits_{u=1}^{d^2}I^s\Big(\rho,\sum\limits_{i=1}^N\mathbb{G}_i^{(u)}\Big)\leq
\sum\limits_{u=1}^{d^2}V\Big(\rho,\sum\limits_{i=1}^N\mathbb{G}_i^{(u)}\Big)$, then quantum state $\rho$ is entangled and can be  detected by entanglement
criteria of Theorem 2 based on the generalized Wigner-Yanase skew information for $-1\leq s\leq0$.

(B3) If $\sum\limits_{u=1}^{d^2}I^s\Big(\rho,\sum\limits_{i=1}^N\mathbb{G}_i^{(u)}\Big)
\leq Nd\eta-N\dfrac{\eta d^2+1}{d(d+1)}\leq
\sum\limits_{u=1}^{d^2}V\Big(\rho,\sum\limits_{i=1}^N\mathbb{G}_i^{(u)}\Big)$,
 then neither the
entanglement criteria of Theorem 2 based on  the generalized Wigner-Yanase skew information for $-1\leq s\leq0$ nor that based on variance  can confirm whether or not $\rho$ is entangled.

From (B1) and (B2), we observe the complementarity of inequalities (\ref{GSIC-SI}) and (\ref{GSIC-SII}), that is, inequality (\ref{GSIC-SI}) can ascertain some entangled states that cannot be detected by inequality (\ref{GSIC-SII}), and vice versa. Next, we will utilize specific examples to demonstrate the complementary relationship between inequality (\ref{GSIC-SI}) and inequality (\ref{GSIC-SII}).

$\emph{Example 3}.$ Consider the family $N$-qubit states given by a
mixture of white noise and the $W$ state $$\rho_3(p)=p|W_N\rangle\langle W_N|+\dfrac{1-p}{2^N}\textbf{1},$$
where $N\geq3$ and $$|W_N\rangle=\dfrac{|10\cdots0\rangle+|01\cdots0\rangle+\cdots+|0\cdots01\rangle}{\sqrt{N}}.$$

For our Theorem 2, let $s=-1$, $\eta=\frac{1}{4}$ and GSICs are given in Appendix D for $d=2$.
For $\rho_3(p)$, we find that the  parameter ranges of entangled states detected by inequality (\ref{GSIC-SI}) and Proposition 1 of Ref. \cite{HongLuoSong2015} are both
 $\frac{N(2^{N-1}-1)+\sqrt{N^2(2^{N-1}-1)^2+2^{N+1}N(3N-2)}}{2^N(3N-2)}<p\leq1$,
  while  inequality (\ref{GSIC-SII}) is unable to identify any entanglement.
The  parameter ranges of entangled states detected by inequality (\ref{GSIC-SI}) and Proposition 1 of Ref. \cite{HongLuoSong2015} for $N=3,4,\cdots,7$ are displayed in Table II.
\begin{table}
\caption{\label{tab:table1}
For $\rho_3(p)=p|W_N\rangle\langle W_N|+\dfrac{1-p}{2^N}\textbf{1}$, the parameter ranges of entangled states detected by inequality (\ref{GSIC-SI}) and Proposition 1 of Ref. \cite{HongLuoSong2015} are same.
When $p_N<p\leq1$,
$\rho_3(p)$ is entangled according to inequality (\ref{GSIC-SI}) and Proposition 1 of Ref. \cite{HongLuoSong2015} for $N=3,4,\cdots,7$.}
\begin{ruledtabular}
\begin{tabular}{ccccccc}
\diagbox [width=2em,trim=l]{}{$N$}&3&4&5&6&7\\
\hline\\
$p_N$& $0.5254$  & $0.4589$ &  $0.4181$  &  $0.3931$ &
$0.3779$ \\
\end{tabular}
\end{ruledtabular}
\end{table}
\begin{table}
\end{table}

$\emph{Example 4}.$  Consider a family of two-qutrit quantum states in the form $$\rho_4(p)=p|\psi\rangle\langle\psi|+\dfrac{1-p}{3^2}\textbf{1},$$
where $|\psi\rangle=\dfrac{|01\rangle-|10\rangle+|02\rangle-|20\rangle+|12\rangle-|21\rangle}{\sqrt{6}}$.

For our Theorem 2, let $s=-1$, $\eta=\frac{1}{9}$ and GSICs are given in Appendix D for $d=3$.
By calculations, inequality (\ref{GSIC-SII}) shows that  $\rho_4(p)$ is an entangled state if $p>0.4495$.
It is worth noting that inequality (\ref{GSIC-SI}) and Proposition 1 of Ref. \cite{HongLuoSong2015} cannot detect  entangled states, but inequality (\ref{GSIC-SII}) can detect some entangled states in this case.

Example 3 and Example 4 illustrate that inequality (\ref{GSIC-SI}) based on the generalized Wigner-Yanase skew information for $-1\leq s\leq0$ and inequality (\ref{GSIC-SII}) based on variance, have varying abilities to detect entanglement in different quantum states, and they are complementary to each other. The combination of
the two criteria based on inequality (\ref{GSIC-SI})  and inequality (\ref{GSIC-SII}) have a more powerful  ability of
entanglement detection. Clearly, the combination of these two entanglement criteria can detected some entangled states undetected by Proposition 1 of Ref. \cite{HongLuoSong2015}.

\section{Disscussion}
In this paper, we introduce several families of separability criteria for detecting the entanglement of multipartite or high-dimensional quantum systems via the generalized Wigner-Yanase skew information $I_s(\rho,X)$ for $-1\leq s\leq0$ and variance. These criteria are classified into two categories based on the observables involved in the generalized Wigner-Yanase skew information or variance. The first is based on mutually unbiased measurements, and the second on general symmetric informationally complete measures. For both the first and the second category, we thoroughly illustrate the complementary relationship between entanglement criteria based on the generalized skew information for $-1\leq s\leq0$ and variance, and demonstrate their individual strengths through concrete examples. This further underscores the superior entanglement detection capability achieved through the combination of these two criteria.

In multipartite quantum systems, there exist various definitions to characterize entanglement,  thus we can endeavor to develop corresponding criteria for detecting different types of entanglement through the methods we propose. This will facilitate our investigation into the complex structures of multipartite entanglement.
Additionally, even though we have obtained entanglement criteria based on the generalized skew information and variance, we still
need to explore criteria derived from different tools and approaches, such as constructing relevant matrices from the perspective of the generalized Wigner-Yanase skew information.

\begin{center}
{\bf ACKNOWLEDGMENTS}
\end{center}

This work was supported by  the National Natural Science Foundation of China under Grant No. 11701135, the Hebei
Natural Science Foundation of China under Grant No. A2017403025, supported by National Pre-research Funds of Hebei GEO University in 2023 (Grant KY202316), PhD Research Startup Foundation of Hebei GEO University (Grant BQ201615).

\appendix

\section{The proof of Theorem 1 }

In an $N$-partite quantum system  $\mathcal{H}_1\otimes \mathcal{H}_2\otimes\cdots \otimes \mathcal{H}_N$ with $\textrm{dim}(\mathcal{H}_i) = d$,
if the pure  state  $|\Psi\rangle$ is fully separable,
it can be written as $|\Psi\rangle=\bigotimes\limits_{i=1}^N|\psi_i\rangle$, where $|\psi_i\rangle$ is the substate of subsystem $\mathcal{H}_i$.
Then we have
\begin{align}
&\sum\limits_{u=1}^{d+1}\sum\limits_{v=1}^dV\Big(|\Psi\rangle,\sum\limits_{i=1}^N\mathbb{M}_i^{(uv)}\Big)\label{}\nonumber\\
=&\sum\limits_{u=1}^{d+1}\sum\limits_{v=1}^d\sum\limits_{i=1}^{N}V\Big(|\psi_i\rangle,M_i^{(uv)}\Big)\label{V0}\\
=&\sum\limits_{i=1}^{N}\Big\{\sum\limits_{u=1}^{d+1}\sum\limits_{v=1}^d\textrm{Tr}\Big[(M_i^{(uv)})^2|\psi_i\rangle\langle\psi_i|\Big]\label{}\nonumber\\
&-\sum\limits_{u=1}^{d+1}\sum\limits_{v=1}^d\Big[\textrm{Tr}(M_{i}^{(uv)}|\psi_i\rangle\langle\psi_i|)\Big]^2\Big\}\label{V1}\\
=&N\kappa d-N.\label{V2}
\end{align}
Here Eqs. (\ref{V0}) and  (\ref{V1}) hold by use of  the additivity and the definition of the variance, respectively.
Eq. (\ref{V2}) is true because of the equality (that is,
$\sum\limits_{u=1}^{d+1}\sum\limits_{v=1}^d(M_{i}^{(uv)})^2
=\kappa(d+1) \textbf{1}_i$ \cite{LaiLuoPRA2022})
and
the equality (that is,
$\sum\limits_{u=1}^{d+1}\sum\limits_{v=1}^d\langle\psi_i|M_{i}^{(uv)}|\psi_i\rangle^2=1+\kappa$ \cite{KalevGour2014,LaiLuoPRA2022,Rastegin2015}).

Consider any $N$-partite fully separable mixed state $\rho=\sum\limits_lp_l|\Psi^{(l)}\rangle\langle\Psi^{(l)}|$ in Hilbert space $\mathcal{H}_1\otimes \mathcal{H}_2\otimes\cdots \otimes \mathcal{H}_N$ with $dim(\mathcal{H}_i) = d,$ where pure state $|\Psi^{(l)}\rangle$ is fully separable.  When  $-1\leq s\leq0$, we can get
\begin{equation*}
\begin{aligned}
&\sum\limits_{u=1}^{d+1}\sum\limits_{v=1}^dI^s\Big(\rho,\sum\limits_{i=1}^N\mathbb{M}_i^{(uv)}\Big)\\
\leq&\sum\limits_lp_l\sum\limits_{u=1}^{d+1}\sum\limits_{v=1}^dI^s\Big(|\Psi^{(l)}\rangle,\sum\limits_{i=1}^N\mathbb{M}_i^{(uv)}\Big)\\
=&\sum\limits_lp_l\sum\limits_{u=1}^{d+1}\sum\limits_{v=1}^dV\Big(|\Psi^{(l)}\rangle,\sum\limits_{i=1}^N\mathbb{M}_i^{(uv)}\Big)\\
= & N\kappa d-N,
\end{aligned}
\end{equation*}
   by the convexity of  the generalized Wigner-Yanase skew information for $-1\leq s\leq0$, equivalence between variance and  the generalized Wigner-Yanase skew information for pure states, and Eq.  (\ref{V2}).
We can also obtain
\begin{equation*}
\begin{aligned}
&\sum\limits_{u=1}^{d+1}\sum\limits_{v=1}^dV\Big(\rho,\sum\limits_{i=1}^N\mathbb{M}_i^{(uv)}\Big)\\
\geq&\sum\limits_lp_l\sum\limits_{u=1}^{d+1}\sum\limits_{v=1}^dV\Big(|\Psi^{(l)}\rangle,\sum\limits_{i=1}^N\mathbb{M}_i^{(uv)}\Big)\\
= &N\kappa d-N,
\end{aligned}
\end{equation*}
by the concave of the variance and Eq. (\ref{V2}). So far, we have demonstrated that inequalities (\ref{MUMs-SI})  and (\ref{MUMs-SII}) are valid for fully separable states $\rho$. From the above proof, we also observe that both inequality (\ref{MUMs-SI}) and inequality (\ref{MUMs-SII}) hold with equality for fully separable pure states.

\section{The proof of Theorem 2 }
 For an $N$-partite quantum system $\mathcal{H}_1\otimes \mathcal{H}_2\otimes\cdots \otimes \mathcal{H}_N$ with $\textrm{dim}(\mathcal{H}_i) = d$, if a pure  state  $|\Psi\rangle$ is fully separable, it can be represented as
$|\Psi\rangle=\bigotimes\limits_{i=1}^N|\psi_{i}\rangle$. Then we can obtain
\begin{align}
&\sum\limits_{u=1}^{d^2}V\Big(|\Psi\rangle,\sum\limits_{i=1}^N\mathbb{G}_i^{(u)}\Big)\label{}\nonumber\\
=&\sum\limits_{u=1}^{d^2}\sum\limits_{i=1}^{N}\Big\{\textrm{Tr}\Big[(G^{(u)}_i)^2|\psi_i\rangle\langle\psi_i|\Big]
-\Big[\textrm{Tr}(G^{(u)}_{i}|\psi_{i}\rangle\langle\psi_{i}|)\Big]^2\Big\}\label{}\nonumber\\
=&Nd\eta-N\dfrac{\eta d^2+1}{d(d+1)}.\label{vG1}
\end{align}
Here Eq.  (\ref{vG1}) holds because of the equality
$\sum\limits_{u=1}^{d^2}(G^{(u)}_{i})^2= d\eta \textbf{1}_i$ \cite{LaiLuoPRA2022},
and
the equality
$\sum\limits_{u=1}^{d^2}\langle\psi_{i}|G_{i}^{(u)}|\psi_{i}\rangle^2=\dfrac{\eta d^2+1}{d(d+1)}$ \cite{LaiLuoPRA2022,Rastegin2014}.

Suppose  $\rho=\sum\limits_lp_l|\Psi^{(l)}\rangle\langle\Psi^{(l)}|$ in Hilbert space $\mathcal{H}_1\otimes \mathcal{H}_2\otimes\cdots \otimes \mathcal{H}_N$ with $\textrm{dim}(\mathcal{H}_i) = d$ be any $N$-partite fully separable mixed state, where pure state $|\Psi^{(l)}\rangle$ is fully separable.
Using the convexity of  the generalized Wigner-Yanase skew information for $-1\leq s\leq0$, equivalence between variance and the  generalized Wigner-Yanase skew information for pure states, and Eq.  (\ref{vG1}),
one has
\begin{equation*}
\begin{aligned}
&\sum\limits_{u=1}^{d^2}I^s\Big(\rho,\sum\limits_{i=1}^N\mathbb{G}_i^{(u)}\Big)\\
\leq&\sum\limits_lp_l\sum\limits_{u=1}^{d^2}I^s\Big(|\Psi^{(l)}\rangle,\sum\limits_{i=1}^N\mathbb{G}_i^{(u)}\Big)\\
=&\sum\limits_lp_l\sum\limits_{u=1}^{d^2}V\Big(|\Psi^{(l)}\rangle,\sum\limits_{i=1}^N\mathbb{G}_i^{(u)}\Big)\\
= & Nd\eta-N\dfrac{\eta d^2+1}{d(d+1)}
\end{aligned}
\end{equation*}
when $-1\leq s\leq0$;
using the concave of the variance and Eq.  (\ref{vG1}), one has
\begin{equation*}
\begin{aligned}
&\sum\limits_{u=1}^{d^2}V\Big(\rho,\sum\limits_{i=1}^N\mathbb{G}_i^{(u)}\Big)\\
\geq&\sum\limits_lp_l\sum\limits_{u=1}^{d^2}V\Big(|\Psi^{(l)}\rangle,\sum\limits_{i=1}^N\mathbb{G}_i^{(u)}\Big)\\
= &Nd\eta-N\dfrac{\eta d^2+1}{d(d+1)}.
\end{aligned}
\end{equation*}
Therefore, for any fully separable state $\rho$, both inequalities (\ref{GSIC-SI}) and (\ref{GSIC-SII}) are valid.
And when $\rho$ is a fully separable pure state, the equality holds in both of these inequalities.

\section{The concrete MUMs  \cite{WiesniakPaterekZeilinger2011,DurtBengtssonZyczkowski2010}}
$\mathcal{M}^{(u)}=\{M^{(uv)}: v=1,2,\cdots,d\}$, with $M^{(uv)}=|\varphi^{(uv)}\rangle\langle\varphi^{(uv)}|$ and $u=1,2,\cdots,d+1.$
Let $a_m=e^{\frac{2\pi i}{m}}.$
$$1.\quad d=2$$
\begin{equation*}
\begin{aligned}
&|\varphi^{(11)}\rangle=|0\rangle,\quad |\varphi^{(12)}\rangle=|1\rangle,\\
&|\varphi^{(21)}\rangle=\frac{|0\rangle+|1\rangle}{\sqrt{2}},\quad |\varphi^{(22)}\rangle=\frac{|0\rangle-|1\rangle}{\sqrt{2}},\\
&|\varphi^{(31)}\rangle=\frac{|0\rangle+i|1\rangle}{\sqrt{2}},\quad |\varphi^{(32)}\rangle=\frac{|0\rangle-i|1\rangle}{\sqrt{2}}
\end{aligned}
\end{equation*}
$$2.\quad d=3 $$
\begin{equation*}
\begin{aligned}
&|\varphi^{(11)}\rangle=|0\rangle,\quad |\varphi^{(12)}\rangle=|1\rangle, \quad|\varphi^{(13)}\rangle=|2\rangle,\\
&|\varphi^{(21)}\rangle=\frac{|0\rangle+|1\rangle+|2\rangle}{\sqrt{3}}, \quad|\varphi^{(22)}\rangle=\frac{|0\rangle+a_3|1\rangle+a_3^2|2\rangle}{\sqrt{3}},\\
&|\varphi^{(23)}\rangle=\frac{|0\rangle+a_3^2|1\rangle+a_3|2\rangle}{\sqrt{3}},\quad
|\varphi^{(31)}\rangle=\frac{|0\rangle+a_3|1\rangle+a_3|2\rangle}{\sqrt{3}},\\
\end{aligned}
\end{equation*}
\begin{equation*}
\begin{aligned}
& |\varphi^{(32)}\rangle=\frac{|0\rangle+a_3^2|1\rangle+|2\rangle}{\sqrt{3}},\quad
|\varphi^{(33)}\rangle=\frac{|0\rangle+|1\rangle+a_3^2|2\rangle}{\sqrt{3}},\\
&|\varphi^{(41)}\rangle=\frac{|0\rangle+a_3^2|1\rangle+a_3^2|2\rangle}{\sqrt{3}}, \quad |\varphi^{(42)}\rangle=\frac{|0\rangle+|1\rangle+a_3|2\rangle}{\sqrt{3}},\\
&|\varphi^{(43)}\rangle=\frac{|0\rangle+a_3|1\rangle+|2\rangle}{\sqrt{3}}.
\end{aligned}
\end{equation*}

$$3.\quad d=4$$
\begin{equation*}
\begin{aligned}
&|\varphi^{(11)}\rangle=|0\rangle,\quad|\varphi^{(12)}\rangle=|1\rangle,\\
&|\varphi^{(13)}\rangle=|2\rangle, \quad|\varphi^{(14)}\rangle=|3\rangle,\\
&|\varphi^{(21)}\rangle=\frac{|0\rangle+|1\rangle+|2\rangle+|3\rangle}{2},\\
&|\varphi^{(22)}\rangle=\frac{|0\rangle-|1\rangle+|2\rangle-|3\rangle}{2},\\
&|\varphi^{(23)}\rangle=\frac{|0\rangle+|1\rangle-|2\rangle-|3\rangle}{2},\\
&|\varphi^{(24)}\rangle=\frac{|0\rangle-|1\rangle-|2\rangle+|3\rangle}{2},\\
&|\varphi^{(31)}\rangle=\frac{|0\rangle+i|1\rangle+i|2\rangle-|3\rangle}{2},\\
&|\varphi^{(32)}\rangle=\frac{|0\rangle-i|1\rangle+i|2\rangle+|3\rangle}{2},\\
&|\varphi^{(33)}\rangle=\frac{|0\rangle+i|1\rangle-i|2\rangle+|3\rangle}{2},\\
&|\varphi^{(34)}\rangle=\frac{|0\rangle-i|1\rangle-i|2\rangle-|3\rangle}{2},\\
&|\varphi^{(41)}\rangle=\frac{|0\rangle+i|1\rangle+|2\rangle-i|3\rangle}{2},\\
&|\varphi^{(42)}\rangle=\frac{|0\rangle-i|1\rangle+|2\rangle+i|3\rangle}{2},\\
&|\varphi^{(43)}\rangle=\frac{|0\rangle+i|1\rangle-|2\rangle+i|3\rangle}{2},\\
&|\varphi^{(44)}\rangle=\frac{|0\rangle-i|1\rangle-|2\rangle-i|3\rangle}{2},\\
&|\varphi^{(51)}\rangle=\frac{|0\rangle+|1\rangle+i|2\rangle-i|3\rangle}{2},\\
&|\varphi^{(52)}\rangle=\frac{|0\rangle-|1\rangle+i|2\rangle+i|3\rangle}{2},\\
&|\varphi^{(53)}\rangle=\frac{|0\rangle+|1\rangle-i|2\rangle+i|3\rangle}{2},\\
&|\varphi^{(54)}\rangle=\frac{|0\rangle-|1\rangle-i|2\rangle-i|3\rangle}{2}.
\end{aligned}
\end{equation*}

$$4.\quad d=5$$
\begin{equation*}
\begin{aligned}
&|\varphi^{(11)}\rangle=|0\rangle,\quad|\varphi^{(12)}\rangle=|1\rangle,\\
&|\varphi^{(13)}\rangle=|2\rangle,\quad|\varphi^{(14)}\rangle=|3\rangle,\quad |\varphi^{(15)}\rangle=|4\rangle,\\
&|\varphi^{(21)}\rangle=\frac{|0\rangle+|1\rangle+|2\rangle+|3\rangle+|4\rangle}{\sqrt{5}},\\
\end{aligned}
\end{equation*}
\begin{equation*}
\begin{aligned}
&|\varphi^{(22)}\rangle=\frac{|0\rangle+a_5|1\rangle+a_5^2|2\rangle+a_5^3|3\rangle+a_5^4|4\rangle}{\sqrt{5}},\\
&|\varphi^{(23)}\rangle=\frac{|0\rangle+a_5^2|1\rangle+a_5^4|2\rangle+a_5|3\rangle+a_5^3|4\rangle}{\sqrt{5}},\\
&|\varphi^{(24)}\rangle=\frac{|0\rangle+a_5^3|1\rangle+a_5|2\rangle+a_5^4|3\rangle+a_5^2|4\rangle}{\sqrt{5}},\\
&|\varphi^{(25)}\rangle=\frac{|0\rangle+a_5^4|1\rangle+a_5^3|2\rangle+a_5^2|3\rangle+a_5|4\rangle}{\sqrt{5}},\\
&|\varphi^{(31)}\rangle=\frac{|0\rangle+a_5|1\rangle+a_5^4|2\rangle+a_5^4|3\rangle+a_5|4\rangle}{\sqrt{5}},\\
&|\varphi^{(32)}\rangle=\frac{|0\rangle+a_5^2|1\rangle+a_5|2\rangle+a_5^2|3\rangle+|4\rangle}{\sqrt{5}},\\
&|\varphi^{(33)}\rangle=\frac{|0\rangle+a_5^3|1\rangle+a_5^3|2\rangle+|3\rangle+a_5^4|4\rangle}{\sqrt{5}},\\
&|\varphi^{(34)}\rangle=\frac{|0\rangle+a_5^4|1\rangle+|2\rangle+a_5^3|3\rangle+a_5^3|4\rangle}{\sqrt{5}},\\
&|\varphi^{(35)}\rangle=\frac{|0\rangle+|1\rangle+a_5^2|2\rangle+a_5|3\rangle+a_5^2|4\rangle}{\sqrt{5}},\\
&|\varphi^{(41)}\rangle=\frac{|0\rangle+a_5^2|1\rangle+a_5^3|2\rangle+a_5^3|3\rangle+a_5^2|4\rangle}{\sqrt{5}},\\
&|\varphi^{(42)}\rangle=\frac{|0\rangle+a_5^3|1\rangle+|2\rangle+a_5|3\rangle+a_5|4\rangle}{\sqrt{5}},\\
&|\varphi^{(43)}\rangle=\frac{|0\rangle+a_5^4|1\rangle+a_5^2|2\rangle+a_5^4|3\rangle+|4\rangle}{\sqrt{5}},\\
&|\varphi^{(44)}\rangle=\frac{|0\rangle+|1\rangle+a_5^4|2\rangle+a_5^2|3\rangle+a_5^4|4\rangle}{\sqrt{5}},\\
&|\varphi^{(45)}\rangle=\frac{|0\rangle+a_5|1\rangle+a_5|2\rangle+|3\rangle+a_5^3|4\rangle}{\sqrt{5}},\\
&|\varphi^{(51)}\rangle=\frac{|0\rangle+a_5^3|1\rangle+a_5^2|2\rangle+a_5^2|3\rangle+a_5^3|4\rangle}{\sqrt{5}},\\
&|\varphi^{(52)}\rangle=\frac{|0\rangle+a_5^4|1\rangle+a_5^4|2\rangle+|3\rangle+a_5^2|4\rangle}{\sqrt{5}},\\
&|\varphi^{(53)}\rangle=\frac{|0\rangle+|1\rangle+a_5|2\rangle+a_5^3|3\rangle+a_5|4\rangle}{\sqrt{5}},\\
&|\varphi^{(54)}\rangle=\frac{|0\rangle+a_5|1\rangle+a_5^3|2\rangle+a_5|3\rangle+|4\rangle}{\sqrt{5}},\\
&|\varphi^{(55)}\rangle=\frac{|0\rangle+a_5^2|1\rangle+|2\rangle+a_5^4|3\rangle+a_5^4|4\rangle}{\sqrt{5}},\\
&|\varphi^{(61)}\rangle=\frac{|0\rangle+a_5^4|1\rangle+a_5|2\rangle+a_5|3\rangle+a_5^4|4\rangle}{\sqrt{5}},\\
&|\varphi^{(62)}\rangle=\frac{|0\rangle+|1\rangle+a_5^3|2\rangle+a_5^4|3\rangle+a_5^3|4\rangle}{\sqrt{5}},\\
&|\varphi^{(63)}\rangle=\frac{|0\rangle+a_5|1\rangle+|2\rangle+a_5^2|3\rangle+a_5^2|4\rangle}{\sqrt{5}},\\
&|\varphi^{(64)}\rangle=\frac{|0\rangle+a_5^2|1\rangle+a_5^2|2\rangle+|3\rangle+a_5|4\rangle}{\sqrt{5}},\\
&|\varphi^{(65)}\rangle=\frac{|0\rangle+a_5^3|1\rangle+a_5^4|2\rangle+a_5^3|3\rangle+|4\rangle}{\sqrt{5}}.
\end{aligned}
\end{equation*}

\begin{widetext}
$$5.\quad d=8 $$
\begin{equation*}
\begin{aligned}
&|\varphi^{(11)}\rangle=|0\rangle,\quad|\varphi^{(12)}\rangle=|1\rangle,
\quad|\varphi^{(13)}\rangle=|2\rangle,\quad |\varphi^{(14)}\rangle=|3\rangle,\\
&|\varphi^{(15)}\rangle=|4\rangle,\quad|\varphi^{(16)}\rangle=|5\rangle,
\quad|\varphi^{(17)}\rangle=|6\rangle,\quad|\varphi^{(18)}\rangle=|7\rangle,\\
&|\varphi^{(21)}\rangle=\frac{|0\rangle+i|1\rangle+i|2\rangle-|3\rangle+i|4\rangle-|5\rangle-|6\rangle-i|7\rangle}{\sqrt{8}},\\
&|\varphi^{(22)}\rangle=\frac{|0\rangle-i|1\rangle+i|2\rangle+|3\rangle+i|4\rangle+|5\rangle-|6\rangle+i|7\rangle}{\sqrt{8}},\\
&|\varphi^{(23)}\rangle=\frac{|0\rangle+|1\rangle-i|2\rangle+|3\rangle+i|4\rangle-|5\rangle+|6\rangle+i|7\rangle}{\sqrt{8}},\\
&|\varphi^{(24)}\rangle=\frac{|0\rangle-i|1\rangle-i|2\rangle-|3\rangle+i|4\rangle+|5\rangle+|6\rangle-i|7\rangle}{\sqrt{8}},\\
&|\varphi^{(25)}\rangle=\frac{|0\rangle+i|1\rangle+i|2\rangle-|3\rangle-i|4\rangle+|5\rangle+|6\rangle+i|7\rangle}{\sqrt{8}},\\
&|\varphi^{(26)}\rangle=\frac{|0\rangle-i|1\rangle+i|2\rangle+|3\rangle-i|4\rangle-|5\rangle+|6\rangle-i|7\rangle}{\sqrt{8}},\\
&|\varphi^{(27)}\rangle=\frac{|0\rangle+i|1\rangle-i|2\rangle+|3\rangle-i|4\rangle+|5\rangle-|6\rangle-i|7\rangle}{\sqrt{8}},\\
&|\varphi^{(28)}\rangle=\frac{|0\rangle-i|1\rangle-i|2\rangle-|3\rangle-i|4\rangle-|5\rangle-|6\rangle+i|7\rangle}{\sqrt{8}},\\
&|\varphi^{(31)}\rangle=\frac{|0\rangle+|1\rangle+i|2\rangle-i|3\rangle+i|4\rangle+i|5\rangle+|6\rangle-|7\rangle}{\sqrt{8}},\\
&|\varphi^{(32)}\rangle=\frac{|0\rangle-|1\rangle+i|2\rangle+i|3\rangle+i|4\rangle-i|5\rangle+|6\rangle+|7\rangle}{\sqrt{8}},\\
&|\varphi^{(33)}\rangle=\frac{|0\rangle+|1\rangle-i|2\rangle+i|3\rangle+i|4\rangle+i|5\rangle-|6\rangle+|7\rangle}{\sqrt{8}},\\
&|\varphi^{(34)}\rangle=\frac{|0\rangle-|1\rangle-i|2\rangle-i|3\rangle+i|4\rangle-i|5\rangle-|6\rangle-|7\rangle}{\sqrt{8}},\\
&|\varphi^{(35)}\rangle=\frac{|0\rangle+|1\rangle+i|2\rangle-i|3\rangle-i|4\rangle-i|5\rangle-|6\rangle+|7\rangle}{\sqrt{8}},\\
&|\varphi^{(36)}\rangle=\frac{|0\rangle-|1\rangle+i|2\rangle+i|3\rangle-i|4\rangle+i|5\rangle-|6\rangle-|7\rangle}{\sqrt{8}},\\
&|\varphi^{(37)}\rangle=\frac{|0\rangle+|1\rangle-i|2\rangle+i|3\rangle-i|4\rangle-i|5\rangle+|6\rangle-|7\rangle}{\sqrt{8}},\\
&|\varphi^{(38)}\rangle=\frac{|0\rangle-|1\rangle-i|2\rangle-i|3\rangle-i|4\rangle+i|5\rangle+|6\rangle+|7\rangle}{\sqrt{8}},\\
&|\varphi^{(41)}\rangle=\frac{|0\rangle+i|1\rangle+|2\rangle-i|3\rangle+i|4\rangle+|5\rangle-i|6\rangle-|7\rangle}{\sqrt{8}},\\
&|\varphi^{(42)}\rangle=\frac{|0\rangle-i|1\rangle+|2\rangle+i|3\rangle+i|4\rangle-|5\rangle-i|6\rangle+|7\rangle}{\sqrt{8}},\\
&|\varphi^{(43)}\rangle=\frac{|0\rangle+i|1\rangle-|2\rangle+i|3\rangle+i|4\rangle+|5\rangle+i|6\rangle+|7\rangle}{\sqrt{8}},\\
&|\varphi^{(44)}\rangle=\frac{|0\rangle-i|1\rangle-|2\rangle-i|3\rangle+i|4\rangle-|5\rangle+i|6\rangle-|7\rangle}{\sqrt{8}},\\
&|\varphi^{(45)}\rangle=\frac{|0\rangle+i|1\rangle+|2\rangle-i|3\rangle-i|4\rangle-|5\rangle+i|6\rangle+|7\rangle}{\sqrt{8}},\\
&|\varphi^{(46)}\rangle=\frac{|0\rangle-i|1\rangle+|2\rangle+i|3\rangle-i|4\rangle+|5\rangle+i|6\rangle-|7\rangle}{\sqrt{8}},\\
\end{aligned}
\end{equation*}
\begin{equation*}
\begin{aligned}
&|\varphi^{(47)}\rangle=\frac{|0\rangle+i|1\rangle-|2\rangle+i|3\rangle-i|4\rangle-|5\rangle-i|6\rangle-|7\rangle}{\sqrt{8}},\\
&|\varphi^{(48)}\rangle=\frac{|0\rangle-i|1\rangle-|2\rangle-i|3\rangle-i|4\rangle+|5\rangle-i|6\rangle+|7\rangle}{\sqrt{8}},\\
&|\varphi^{(51)}\rangle=\frac{|0\rangle+|1\rangle+|2\rangle-|3\rangle+i|4\rangle-i|5\rangle+i|6\rangle+i|7\rangle}{\sqrt{8}},\\
&|\varphi^{(52)}\rangle=\frac{|0\rangle-|1\rangle+|2\rangle+|3\rangle+i|4\rangle+i|5\rangle+i|6\rangle-i|7\rangle}{\sqrt{8}},\\
&|\varphi^{(53)}\rangle=\frac{|0\rangle+|1\rangle-|2\rangle+|3\rangle+i|4\rangle-i|5\rangle-i|6\rangle-i|7\rangle}{\sqrt{8}},\\
&|\varphi^{(54)}\rangle=\frac{|0\rangle-|1\rangle-|2\rangle-|3\rangle+i|4\rangle+i|5\rangle-i|6\rangle+i|7\rangle}{\sqrt{8}},\\
&|\varphi^{(55)}\rangle=\frac{|0\rangle+|1\rangle+|2\rangle-|3\rangle-i|4\rangle+i|5\rangle-i|6\rangle-i|7\rangle}{\sqrt{8}},\\
&|\varphi^{(56)}\rangle=\frac{|0\rangle-|1\rangle+|2\rangle+|3\rangle-i|4\rangle-i|5\rangle-i|6\rangle+i|7\rangle}{\sqrt{8}},\\
&|\varphi^{(57)}\rangle=\frac{|0\rangle+|1\rangle-|2\rangle+|3\rangle-i|4\rangle+i|5\rangle+i|6\rangle+i|7\rangle}{\sqrt{8}},\\
&|\varphi^{(58)}\rangle=\frac{|0\rangle-|1\rangle-|2\rangle-|3\rangle-i|4\rangle-i|5\rangle+i|6\rangle-i|7\rangle}{\sqrt{8}},\\
&|\varphi^{(61)}\rangle=\frac{|0\rangle+i|1\rangle+i|2\rangle+|3\rangle+|4\rangle-i|5\rangle+i|6\rangle-|7\rangle}{\sqrt{8}},\\
&|\varphi^{(62)}\rangle=\frac{|0\rangle-i|1\rangle+i|2\rangle-|3\rangle+|4\rangle+i|5\rangle+i|6\rangle+|7\rangle}{\sqrt{8}},\\
&|\varphi^{(63)}\rangle=\frac{|0\rangle+i|1\rangle-i|2\rangle-|3\rangle+|4\rangle-i|5\rangle-i|6\rangle+|7\rangle}{\sqrt{8}},\\
&|\varphi^{(64)}\rangle=\frac{|0\rangle-i|1\rangle-i|2\rangle+|3\rangle+|4\rangle+i|5\rangle-i|6\rangle-|7\rangle}{\sqrt{8}},\\
&|\varphi^{(65)}\rangle=\frac{|0\rangle+i|1\rangle+i|2\rangle+|3\rangle-|4\rangle+i|5\rangle-i|6\rangle+|7\rangle}{\sqrt{8}},\\
&|\varphi^{(66)}\rangle=\frac{|0\rangle-i|1\rangle+i|2\rangle-|3\rangle-|4\rangle-i|5\rangle-i|6\rangle-|7\rangle}{\sqrt{8}},\\
&|\varphi^{(67)}\rangle=\frac{|0\rangle+i|1\rangle-i|2\rangle-|3\rangle-|4\rangle+i|5\rangle+i|6\rangle-|7\rangle}{\sqrt{8}},\\
&|\varphi^{(68)}\rangle=\frac{|0\rangle-i|1\rangle-i|2\rangle+|3\rangle-|4\rangle-i|5\rangle+i|6\rangle+|7\rangle}{\sqrt{8}},\\
&|\varphi^{(71)}\rangle=\frac{|0\rangle+|1\rangle+i|2\rangle+i|3\rangle+|4\rangle-|5\rangle-i|6\rangle+i|7\rangle}{\sqrt{8}},\\
&|\varphi^{(72)}\rangle=\frac{|0\rangle-|1\rangle+i|2\rangle-i|3\rangle+|4\rangle+|5\rangle-i|6\rangle-i|7\rangle}{\sqrt{8}},\\
&|\varphi^{(73)}\rangle=\frac{|0\rangle+|1\rangle-i|2\rangle-i|3\rangle+|4\rangle-|5\rangle+i|6\rangle-i|7\rangle}{\sqrt{8}},\\
&|\varphi^{(74)}\rangle=\frac{|0\rangle-|1\rangle-i|2\rangle+i|3\rangle+|4\rangle+|5\rangle+i|6\rangle+i|7\rangle}{\sqrt{8}},\\
&|\varphi^{(75)}\rangle=\frac{|0\rangle+|1\rangle+i|2\rangle+i|3\rangle-|4\rangle+|5\rangle+i|6\rangle-i|7\rangle}{\sqrt{8}},\\
&|\varphi^{(76)}\rangle=\frac{|0\rangle-|1\rangle+i|2\rangle-i|3\rangle-|4\rangle-|5\rangle+i|6\rangle+i|7\rangle}{\sqrt{8}},\\
&|\varphi^{(77)}\rangle=\frac{|0\rangle+|1\rangle-i|2\rangle-i|3\rangle-|4\rangle+|5\rangle-i|6\rangle+i|7\rangle}{\sqrt{8}},\\
&|\varphi^{(78)}\rangle=\frac{|0\rangle-|1\rangle-i|2\rangle+i|3\rangle-|4\rangle-|5\rangle-i|6\rangle-i|7\rangle}{\sqrt{8}},\\
&|\varphi^{(81)}\rangle=\frac{|0\rangle+i|1\rangle+|2\rangle-i|3\rangle+|4\rangle-i|5\rangle-|6\rangle+i|7\rangle}{\sqrt{8}},\\
\end{aligned}
\end{equation*}
\begin{equation*}
\begin{aligned}
&|\varphi^{(82)}\rangle=\frac{|0\rangle-i|1\rangle+|2\rangle+i|3\rangle+|4\rangle+i|5\rangle-|6\rangle-i|7\rangle}{\sqrt{8}},\\
&|\varphi^{(83)}\rangle=\frac{|0\rangle+i|1\rangle-|2\rangle+i|3\rangle+|4\rangle-i|5\rangle+|6\rangle-i|7\rangle}{\sqrt{8}},\\
&|\varphi^{(84)}\rangle=\frac{|0\rangle-i|1\rangle-|2\rangle-i|3\rangle+|4\rangle+i|5\rangle+|6\rangle+i|7\rangle}{\sqrt{8}},\\
&|\varphi^{(85)}\rangle=\frac{|0\rangle+i|1\rangle+|2\rangle-i|3\rangle-|4\rangle+i|5\rangle+|6\rangle-i|7\rangle}{\sqrt{8}},\\
&|\varphi^{(86)}\rangle=\frac{|0\rangle-i|1\rangle+|2\rangle+i|3\rangle-|4\rangle-i|5\rangle+|6\rangle+i|7\rangle}{\sqrt{8}},\\
&|\varphi^{(87)}\rangle=\frac{|0\rangle+i|1\rangle-|2\rangle+i|3\rangle-|4\rangle+i|5\rangle-|6\rangle+i|7\rangle}{\sqrt{8}},\\
&|\varphi^{(88)}\rangle=\frac{|0\rangle-i|1\rangle-|2\rangle-i|3\rangle-|4\rangle-i|5\rangle-|6\rangle-i|7\rangle}{\sqrt{8}},\\
&|\varphi^{(91)}\rangle=\frac{|0\rangle+|1\rangle+|2\rangle+|3\rangle+|4\rangle+|5\rangle+|6\rangle-|7\rangle}{\sqrt{8}},\\
&|\varphi^{(92)}\rangle=\frac{|0\rangle-|1\rangle+|2\rangle-|3\rangle+|4\rangle-|5\rangle+|6\rangle+|7\rangle}{\sqrt{8}},\\
&|\varphi^{(93)}\rangle=\frac{|0\rangle+|1\rangle-|2\rangle-|3\rangle+|4\rangle+|5\rangle-|6\rangle+|7\rangle}{\sqrt{8}},\\
&|\varphi^{(94)}\rangle=\frac{|0\rangle-|1\rangle-|2\rangle+|3\rangle+|4\rangle-|5\rangle-|6\rangle-|7\rangle}{\sqrt{8}},\\
&|\varphi^{(95)}\rangle=\frac{|0\rangle+|1\rangle+|2\rangle+|3\rangle-|4\rangle-|5\rangle-|6\rangle+|7\rangle}{\sqrt{8}},\\
&|\varphi^{(96)}\rangle=\frac{|0\rangle-|1\rangle+|2\rangle-|3\rangle-|4\rangle+|5\rangle-|6\rangle-|7\rangle}{\sqrt{8}},\\
&|\varphi^{(97)}\rangle=\frac{|0\rangle+|1\rangle-|2\rangle-|3\rangle-|4\rangle-|5\rangle+|6\rangle-|7\rangle}{\sqrt{8}},\\
&|\varphi^{(98)}\rangle=\frac{|0\rangle-|1\rangle-|2\rangle+|3\rangle-|4\rangle+|5\rangle+|6\rangle+|7\rangle}{\sqrt{8}}.
\end{aligned}
\end{equation*}
$$6.\quad d=9$$
\begin{equation*}
\begin{aligned}
&|\varphi^{(11)}\rangle=|0\rangle,\quad |\varphi^{(12)}\rangle=|1\rangle,\quad
|\varphi^{(13)}\rangle=|2\rangle, \quad|\varphi^{(14)}\rangle=|3\rangle,\quad
|\varphi^{(15)}\rangle=|4\rangle,\\
&|\varphi^{(16)}\rangle=|5\rangle,\quad|\varphi^{(17)}\rangle=|6\rangle,\quad|\varphi^{(18)}\rangle=|7\rangle,\quad|\varphi^{(19)}\rangle=|8\rangle,\\
&|\varphi^{(21)}\rangle=\frac{|0\rangle+|1\rangle+|2\rangle+|3\rangle+|4\rangle+|5\rangle+|6\rangle+|7\rangle+|8\rangle}{3},\\
&|\varphi^{(22)}\rangle=\frac{|0\rangle+a_3|1\rangle+a_3^2|2\rangle+|3\rangle+a_3|4\rangle+a_3^2|5\rangle+|6\rangle+a_3|7\rangle+a_3^2|8\rangle}{3},\\
&|\varphi^{(23)}\rangle=\frac{|0\rangle+a_3^2|1\rangle+a_3|2\rangle+|3\rangle+a_3^2|4\rangle+a_3|5\rangle+|6\rangle+a_3^2|7\rangle+a_3|8\rangle}{3},\\
&|\varphi^{(24)}\rangle=\frac{|0\rangle+|1\rangle+|2\rangle+a_3|3\rangle+a_3|4\rangle+a_3|5\rangle+a_3^2|6\rangle+a_3^2|7\rangle+a_3^2|8\rangle}{3},\\
&|\varphi^{(25)}\rangle=\frac{|0\rangle+a_3|1\rangle+a_3^2|2\rangle+a_3|3\rangle+a_3^2|4\rangle+|5\rangle+a_3^2|6\rangle+|7\rangle+a_3|8\rangle}{3},\\
&|\varphi^{(26)}\rangle=\frac{|0\rangle+a_3^2|1\rangle+a_3|2\rangle+a_3|3\rangle+|4\rangle+a_3^2|5\rangle+a_3^2|6\rangle+a_3|7\rangle+|8\rangle}{3},\\
&|\varphi^{(27)}\rangle=\frac{|0\rangle+|1\rangle+|2\rangle+a_3^2|3\rangle+a_3^2|4\rangle+a_3^2|5\rangle+a_3|6\rangle+a_3|7\rangle+a_3|8\rangle}{3},\\
&|\varphi^{(28)}\rangle=\frac{|0\rangle+a_3|1\rangle+a_3^2|2\rangle+a_3^2|3\rangle+|4\rangle+a_3|5\rangle+a_3|6\rangle+a_3^2|7\rangle+|8\rangle}{3},\\
\end{aligned}
\end{equation*}
\begin{equation*}
\begin{aligned}
&|\varphi^{(29)}\rangle=\frac{|0\rangle+a_3^2|1\rangle+a_3|2\rangle+a_3^2|3\rangle+a_3|4\rangle+|5\rangle+a_3|6\rangle+|7\rangle+a_3^2|8\rangle}{3},\\
&|\varphi^{(31)}\rangle=\frac{|0\rangle+a_3|1\rangle+a_3|2\rangle+a_3|3\rangle+a_3^2|4\rangle+a_3^2|5\rangle+a_3|6\rangle+a_3^2|7\rangle+a_3^2|8\rangle}{3},\\
&|\varphi^{(32)}\rangle=\frac{|0\rangle+a_3^2|1\rangle+|2\rangle+a_3|3\rangle+|4\rangle+a_3|5\rangle+a_3|6\rangle+|7\rangle+a_3|8\rangle}{3},\\
&|\varphi^{(33)}\rangle=\frac{|0\rangle+|1\rangle+a_3^2|2\rangle+a_3|3\rangle+a_3|4\rangle+|5\rangle+a_3|6\rangle+a_3|7\rangle+|8\rangle}{3},\\
&|\varphi^{(34)}\rangle=\frac{|0\rangle+a_3|1\rangle+a_3|2\rangle+a_3^2|3\rangle+|4\rangle+|5\rangle+|6\rangle+a_3|7\rangle+a_3|8\rangle}{3},\\
&|\varphi^{(35)}\rangle=\frac{|0\rangle+a_3^2|1\rangle+|2\rangle+a_3^2|3\rangle+a_3|4\rangle+a_3^2|5\rangle+|6\rangle+a_3^2|7\rangle+|8\rangle}{3},\\
&|\varphi^{(36)}\rangle=\frac{|0\rangle+|1\rangle+a_3^2|2\rangle+a_3^2|3\rangle+a_3^2|4\rangle+a_3|5\rangle+|6\rangle+|7\rangle+a_3^2|8\rangle}{3},\\
&|\varphi^{(37)}\rangle=\frac{|0\rangle+a_3|1\rangle+a_3|2\rangle+|3\rangle+a_3|4\rangle+a_3|5\rangle+a_3^2|6\rangle+|7\rangle+|8\rangle}{3},\\
&|\varphi^{(38)}\rangle=\frac{|0\rangle+a_3^2|1\rangle+|2\rangle+|3\rangle+a_3^2|4\rangle+|5\rangle+a_3^2|6\rangle+a_3|7\rangle+a_3^2|8\rangle}{3},\\
&|\varphi^{(39)}\rangle=\frac{|0\rangle+|1\rangle+a_3^2|2\rangle+|3\rangle+|4\rangle+a_3^2|5\rangle+a_3^2|6\rangle+a_3^2|7\rangle+a_3|8\rangle}{3},\\
&|\varphi^{(41)}\rangle=\frac{|0\rangle+a_3^2|1\rangle+a_3^2|2\rangle+a_3^2|3\rangle+a_3|4\rangle+a_3|5\rangle+a_3^2|6\rangle+a_3|7\rangle+a_3|8\rangle}{3},\\
&|\varphi^{(42)}\rangle=\frac{|0\rangle+|1\rangle+a_3|2\rangle+a_3^2|3\rangle+a_3^2|4\rangle+|5\rangle+a_3^2|6\rangle+a_3^2|7\rangle+|8\rangle}{3},\\
&|\varphi^{(43)}\rangle=\frac{|0\rangle+a_3|1\rangle+|2\rangle+a_3^2|3\rangle+|4\rangle+a_3^2|5\rangle+a_3^2|6\rangle+|7\rangle+a_3^2|8\rangle}{3},\\
&|\varphi^{(44)}\rangle=\frac{|0\rangle+a_3^2|1\rangle+a_3^2|2\rangle+|3\rangle+a_3^2|4\rangle+a_3^2|5\rangle+a_3|6\rangle+|7\rangle+|8\rangle}{3},\\
&|\varphi^{(45)}\rangle=\frac{|0\rangle+|1\rangle+a_3|2\rangle+|3\rangle+|4\rangle+a_3|5\rangle+a_3|6\rangle+a_3|7\rangle+a_3^2|8\rangle}{3},\\
&|\varphi^{(46)}\rangle=\frac{|0\rangle+a_3|1\rangle+|2\rangle+|3\rangle+a_3|4\rangle+|5\rangle+a_3|6\rangle+a_3^2|7\rangle+a_3|8\rangle}{3},\\
&|\varphi^{(47)}\rangle=\frac{|0\rangle+a_3^2|1\rangle+a_3^2|2\rangle+a_3|3\rangle+|4\rangle+|5\rangle+|6\rangle+a_3^2|7\rangle+a_3^2|8\rangle}{3},\\
&|\varphi^{(48)}\rangle=\frac{|0\rangle+|1\rangle+a_3|2\rangle+a_3|3\rangle+a_3|4\rangle+a_3^2|5\rangle+|6\rangle+|7\rangle+a_3|8\rangle}{3},\\
&|\varphi^{(49)}\rangle=\frac{|0\rangle+a_3|1\rangle+|2\rangle+a_3|3\rangle+a_3^2|4\rangle+a_3|5\rangle+|6\rangle+a_3|7\rangle+|8\rangle}{3},\\
&|\varphi^{(51)}\rangle=\frac{|0\rangle+a_3|1\rangle+a_3|2\rangle+|3\rangle+|4\rangle+a_3^2|5\rangle+|6\rangle+a_3^2|7\rangle+|8\rangle}{3},\\
&|\varphi^{(52)}\rangle=\frac{|0\rangle+a_3^2|1\rangle+|2\rangle+|3\rangle+a_3|4\rangle+a_3|5\rangle+|6\rangle+|7\rangle+a_3^2|8\rangle}{3},\\
&|\varphi^{(53)}\rangle=\frac{|0\rangle+|1\rangle+a_3^2|2\rangle+|3\rangle+a_3^2|4\rangle+|5\rangle+|6\rangle+a_3|7\rangle+a_3|8\rangle}{3},\\
&|\varphi^{(54)}\rangle=\frac{|0\rangle+a_3|1\rangle+a_3|2\rangle+a_3|3\rangle+a_3|4\rangle+|5\rangle+a_3^2|6\rangle+a_3|7\rangle+a_3^2|8\rangle}{3},\\
&|\varphi^{(55)}\rangle=\frac{|0\rangle+|1\rangle+a_3^2|2\rangle+|3\rangle+a_3^2|4\rangle+a_3^2|5\rangle+a_3^2|6\rangle+a_3^2|7\rangle+a_3|8\rangle}{3},\\
&|\varphi^{(56)}\rangle=\frac{|0\rangle+|1\rangle+a_3^2|2\rangle+a_3|3\rangle+|4\rangle+a_3|5\rangle+a_3^2|6\rangle+|7\rangle+|8\rangle}{3},\\
&|\varphi^{(57)}\rangle=\frac{|0\rangle+a_3|1\rangle+a_3|2\rangle+a_3^2|3\rangle+a_3^2|4\rangle+a_3|5\rangle+a_3|6\rangle+|7\rangle+a_3|8\rangle}{3},\\
&|\varphi^{(58)}\rangle=\frac{|0\rangle+a_3^2|1\rangle+|2\rangle+a_3^2|3\rangle+|4\rangle+|5\rangle+a_3|6\rangle+a_3|7\rangle+|8\rangle}{3},\\
&|\varphi^{(59)}\rangle=\frac{|0\rangle+|1\rangle+a_3^2|2\rangle+a_3^2|3\rangle+a_3|4\rangle+a_3^2|5\rangle+a_3|6\rangle+a_3^2|7\rangle+a_3^2|8\rangle}{3},\\
\end{aligned}
\end{equation*}
\begin{equation*}
\begin{aligned}
&|\varphi^{(61)}\rangle=\frac{|0\rangle+a_3^2|1\rangle+a_3^2|2\rangle+|3\rangle+|4\rangle+a_3|5\rangle+|6\rangle+a_3|7\rangle+|8\rangle}{3},\\
&|\varphi^{(62)}\rangle=\frac{|0\rangle+|1\rangle+a_3|2\rangle+|3\rangle+a_3|4\rangle+|5\rangle+|6\rangle+a_3^2|7\rangle+a_3^2|8\rangle}{3},\\
&|\varphi^{(63)}\rangle=\frac{|0\rangle+a_3|1\rangle+|2\rangle+|3\rangle+a_3^2|4\rangle+a_3^2|5\rangle+|6\rangle+|7\rangle+a_3|8\rangle}{3},\\
&|\varphi^{(64)}\rangle=\frac{|0\rangle+a_3^2|1\rangle+a_3^2|2\rangle+a_3|3\rangle+a_3|4\rangle+a_3^2|5\rangle+a_3^2|6\rangle+|7\rangle+a_3^2|8\rangle}{3},\\
&|\varphi^{(65)}\rangle=\frac{|0\rangle+|1\rangle+a_3|2\rangle+a_3|3\rangle+a_3^2|4\rangle+a_3|5\rangle+a_3^2|6\rangle+a_3|7\rangle+a_3|8\rangle}{3},\\
&|\varphi^{(66)}\rangle=\frac{|0\rangle+a_3|1\rangle+|2\rangle+a_3|3\rangle+|4\rangle+|5\rangle+a_3^2|6\rangle+a_3^2|7\rangle+|8\rangle}{3},\\
&|\varphi^{(67)}\rangle=\frac{|0\rangle+a_3^2|1\rangle+a_3^2|2\rangle+a_3^2|3\rangle+a_3^2|4\rangle+|5\rangle+a_3|6\rangle+a_3^2|7\rangle+a_3|8\rangle}{3},\\
&|\varphi^{(68)}\rangle=\frac{|0\rangle+|1\rangle+a_3|2\rangle+a_3^2|3\rangle+|4\rangle+a_3^2|5\rangle+a_3|6\rangle+|7\rangle+|8\rangle}{3},\\
&|\varphi^{(69)}\rangle=\frac{|0\rangle+a_3|1\rangle+|2\rangle+a_3^2|3\rangle+a_3|4\rangle+a_3|5\rangle+a_3|6\rangle+a_3|7\rangle+a_3^2|8\rangle}{3},\\
&|\varphi^{(71)}\rangle=\frac{|0\rangle+|1\rangle+|2\rangle+a_3|3\rangle+a_3^2|4\rangle+|5\rangle+a_3|6\rangle+|7\rangle+a_3^2|8\rangle}{3},\\
&|\varphi^{(72)}\rangle=\frac{|0\rangle+a_3|1\rangle+a_3^2|2\rangle+a_3|3\rangle+|4\rangle+a_3^2|5\rangle+a_3|6\rangle+a_3|7\rangle+a_3|8\rangle}{3},\\
&|\varphi^{(73)}\rangle=\frac{|0\rangle+a_3^2|1\rangle+a_3|2\rangle+a_3|3\rangle+a_3|4\rangle+a_3|5\rangle+a_3|6\rangle+a_3^2|7\rangle+|8\rangle}{3},\\
&|\varphi^{(74)}\rangle=\frac{|0\rangle+|1\rangle+|2\rangle+a_3^2|3\rangle+|4\rangle+a_3|5\rangle+|6\rangle+a_3^2|7\rangle+a_3|8\rangle}{3},\\
&|\varphi^{(75)}\rangle=\frac{|0\rangle+a_3|1\rangle+a_3^2|2\rangle+a_3^2|3\rangle+a_3|4\rangle+|5\rangle+|6\rangle+|7\rangle+|8\rangle}{3},\\
&|\varphi^{(76)}\rangle=\frac{|0\rangle+a_3^2|1\rangle+a_3|2\rangle+a_3^2|3\rangle+a_3^2|4\rangle+a_3^2|5\rangle+|6\rangle+a_3|7\rangle+a_3^2|8\rangle}{3},\\
&|\varphi^{(77)}\rangle=\frac{|0\rangle+|1\rangle+|2\rangle+|3\rangle+a_3|4\rangle+a_3^2|5\rangle+a_3^2|6\rangle+a_3|7\rangle+|8\rangle}{3},\\
&|\varphi^{(78)}\rangle=\frac{|0\rangle+a_3|1\rangle+a_3^2|2\rangle+|3\rangle+a_3^2|4\rangle+a_3|5\rangle+a_3^2|6\rangle+a_3^2|7\rangle+a_3^2|8\rangle}{3},\\
&|\varphi^{(79)}\rangle=\frac{|0\rangle+a_3^2|1\rangle+a_3|2\rangle+|3\rangle+|4\rangle+|5\rangle+a_3^2|6\rangle+|7\rangle+a_3|8\rangle}{3},\\
&|\varphi^{(81)}\rangle=\frac{|0\rangle+a_3^2|1\rangle+a_3^2|2\rangle+a_3|3\rangle+a_3^2|4\rangle+a_3|5\rangle+a_3|6\rangle+a_3|7\rangle+a_3^2|8\rangle}{3},\\
&|\varphi^{(82)}\rangle=\frac{|0\rangle+|1\rangle+a_3|2\rangle+a_3|3\rangle+|4\rangle+|5\rangle+a_3|6\rangle+a_3^2|7\rangle+a_3|8\rangle}{3},\\
&|\varphi^{(83)}\rangle=\frac{|0\rangle+a_3|1\rangle+|2\rangle+a_3|3\rangle+a_3|4\rangle+a_3^2|5\rangle+a_3|6\rangle+|7\rangle+|8\rangle}{3},\\
&|\varphi^{(84)}\rangle=\frac{|0\rangle+a_3^2|1\rangle+a_3^2|2\rangle+a_3^2|3\rangle+|4\rangle+a_3^2|5\rangle+|6\rangle+|7\rangle+a_3|8\rangle}{3},\\
&|\varphi^{(85)}\rangle=\frac{|0\rangle+|1\rangle+a_3|2\rangle+a_3^2|3\rangle+a_3|4\rangle+a_3|5\rangle+|6\rangle+a_3|7\rangle+|8\rangle}{3},\\
&|\varphi^{(86)}\rangle=\frac{|0\rangle+a_3|1\rangle+|2\rangle+a_3^2|3\rangle+a_3^2|4\rangle+|5\rangle+|6\rangle+a_3^2|7\rangle+a_3^2|8\rangle}{3},\\
&|\varphi^{(87)}\rangle=\frac{|0\rangle+a_3^2|1\rangle+a_3^2|2\rangle+|3\rangle+a_3|4\rangle+|5\rangle+a_3^2|6\rangle+a_3^2|7\rangle+|8\rangle}{3},\\
&|\varphi^{(88)}\rangle=\frac{|0\rangle+|1\rangle+a_3|2\rangle+|3\rangle+a_3^2|4\rangle+a_3^2|5\rangle+a_3^2|6\rangle+|7\rangle+a_3^2|8\rangle}{3},\\
&|\varphi^{(89)}\rangle=\frac{|0\rangle+a_3|1\rangle+|2\rangle+|3\rangle+|4\rangle+a_3|5\rangle+a_3^2|6\rangle+a_3|7\rangle+a_3|8\rangle}{3},\\
&|\varphi^{(91)}\rangle=\frac{|0\rangle+|1\rangle+|2\rangle+a_3^2|3\rangle+a_3|4\rangle+|5\rangle+a_3^2|6\rangle+|7\rangle+a_3|8\rangle}{3},\\
\end{aligned}
\end{equation*}
\begin{equation*}
\begin{aligned}
&|\varphi^{(92)}\rangle=\frac{|0\rangle+a_3|1\rangle+a_3^2|2\rangle+a_3^2|3\rangle+a_3^2|4\rangle+a_3^2|5\rangle+a_3^2|6\rangle+a_3|7\rangle+|8\rangle}{3},\\
&|\varphi^{(93)}\rangle=\frac{|0\rangle+a_3^2|1\rangle+a_3|2\rangle+a_3^2|3\rangle+|4\rangle+a_3|5\rangle+a_3^2|6\rangle+a_3^2|7\rangle+a_3^2|8\rangle}{3},\\
&|\varphi^{(94)}\rangle=\frac{|0\rangle+|1\rangle+|2\rangle+|3\rangle+a_3^2|4\rangle+a_3|5\rangle+a_3|6\rangle+a_3^2|7\rangle+|8\rangle}{3},\\
&|\varphi^{(95)}\rangle=\frac{|0\rangle+a_3|1\rangle+a_3^2|2\rangle+|3\rangle+|4\rangle+|5\rangle+a_3|6\rangle+|7\rangle+a_3^2|8\rangle}{3},\\
&|\varphi^{(96)}\rangle=\frac{|0\rangle+a_3^2|1\rangle+a_3|2\rangle+|3\rangle+a_3|4\rangle+a_3^2|5\rangle+a_3|6\rangle+a_3|7\rangle+a_3|8\rangle}{3},\\
&|\varphi^{(97)}\rangle=\frac{|0\rangle+|1\rangle+|2\rangle+a_3|3\rangle+|4\rangle+a_3^2|5\rangle+|6\rangle+a_3|7\rangle+a_3^2|8\rangle}{3},\\
&|\varphi^{(98)}\rangle=\frac{|0\rangle+a_3|1\rangle+a_3^2|2\rangle+a_3|3\rangle+a_3|4\rangle+a_3|5\rangle+|6\rangle+a_3^2|7\rangle+a_3|8\rangle}{3},\\
&|\varphi^{(99)}\rangle=\frac{|0\rangle+a_3^2|1\rangle+a_3|2\rangle+a_3|3\rangle+a_3^2|4\rangle+|5\rangle+|6\rangle+|7\rangle+|8\rangle}{3},\\
&|\varphi^{(101)}\rangle=\frac{|0\rangle+a_3|1\rangle+a_3|2\rangle+a_3^2|3\rangle+a_3|4\rangle+a_3^2|5\rangle+a_3^2|6\rangle+a_3^2|7\rangle+a_3|8\rangle}{3},\\
&|\varphi^{(102)}\rangle=\frac{|0\rangle+a_3^2|1\rangle+|2\rangle+a_3^2|3\rangle+a_3^2|4\rangle+a_3|5\rangle+a_3^2|6\rangle+|7\rangle+|8\rangle}{3},\\
&|\varphi^{(103)}\rangle=\frac{|0\rangle+|1\rangle+a_3^2|2\rangle+a_3^2|3\rangle+|4\rangle+|5\rangle+a_3^2|6\rangle+a_3|7\rangle+a_3^2|8\rangle}{3},\\
&|\varphi^{(104)}\rangle=\frac{|0\rangle+a_3|1\rangle+a_3|2\rangle+|3\rangle+a_3^2|4\rangle+|5\rangle+a_3|6\rangle+a_3|7\rangle+|8\rangle}{3},\\
&|\varphi^{(105)}\rangle=\frac{|0\rangle+a_3^2|1\rangle+|2\rangle+|3\rangle+|4\rangle+a_3^2|5\rangle+a_3|6\rangle+a_3^2|7\rangle+a_3^2|8\rangle}{3},\\
&|\varphi^{(106)}\rangle=\frac{|0\rangle+|1\rangle+a_3^2|2\rangle+|3\rangle+a_3|4\rangle+a_3|5\rangle+a_3|6\rangle+|7\rangle+a_3|8\rangle}{3},\\
&|\varphi^{(107)}\rangle=\frac{|0\rangle+a_3|1\rangle+a_3|2\rangle+a_3|3\rangle+|4\rangle+a_3|5\rangle+|6\rangle+|7\rangle+a_3^2|8\rangle}{3},\\
&|\varphi^{(108)}\rangle=\frac{|0\rangle+a_3^2|1\rangle+|2\rangle+a_3|3\rangle+a_3|4\rangle+|5\rangle+|6\rangle+a_3|7\rangle+a_3|8\rangle}{3},\\
&|\varphi^{(109)}\rangle=\frac{|0\rangle+|1\rangle+a_3^2|2\rangle+a_3|3\rangle+a_3^2|4\rangle+a_3^2|5\rangle+|6\rangle+a_3^2|7\rangle+|8\rangle}{3}.
\end{aligned}
\end{equation*}

\section{The concrete GSICs  \cite{GourKalev2014,LaiLuoPRA2022,RenesKohoutScottCaves2004}}
$$1. \quad d=2$$
$$G^{(1)}=\frac{1}{12\sqrt{3}}
\left[{
  \begin{array}{cc}
    3\sqrt{3}+1 & -5+i \\
   -5-i & 3\sqrt{3}-1 \\
  \end{array}}
\right],
\quad G^{(2)}=\frac{1}{12\sqrt{3}}
\left[{
  \begin{array}{cc}
    3\sqrt{3}+1 & 1-5i \\
   1+5i & 3\sqrt{3}-1 \\
  \end{array}}
\right],$$
$$G^{(3)}=\frac{1}{12\sqrt{3}}
\left[{
  \begin{array}{cc}
    3\sqrt{3}-5 & 1+i \\
   1-i & 3\sqrt{3}+5 \\
  \end{array}}
\right],
\quad G^{(4)}=\frac{1}{4\sqrt{3}}
\left[{
  \begin{array}{cc}
    \sqrt{3}+1 & 1+i \\
   1-i & \sqrt{3}-1 \\
  \end{array}}
\right].$$

$$2. \quad d=3 $$
$G^{(u)}=\dfrac{1}{3}|\varphi_u\rangle\langle\varphi_u|,$ $u=1,2,\cdots,9.$  Let $b=e^{\frac{2\pi i}{3}}.$
\begin{equation*}
\begin{aligned}
&|\varphi_1\rangle=\frac{1}{\sqrt{2}}(|1\rangle-|2\rangle),\quad |\varphi_2\rangle=\frac{1}{\sqrt{2}}(b|1\rangle-b^2|2\rangle),\quad
|\varphi_3\rangle=\frac{1}{\sqrt{2}}(b^2|1\rangle-b|2\rangle),\\
&|\varphi_4\rangle=\frac{1}{\sqrt{2}}(|0\rangle-|1\rangle),\quad |\varphi_5\rangle=\frac{1}{\sqrt{2}}(b|0\rangle-b^2|1\rangle),\quad
|\varphi_6\rangle=\frac{1}{\sqrt{2}}(b^2|0\rangle-b|1\rangle),\\
\end{aligned}
\end{equation*}
\begin{equation*}
\begin{aligned}
&|\varphi_7\rangle=\frac{1}{\sqrt{2}}(|2\rangle-|0\rangle),\quad |\varphi_8\rangle=\frac{1}{\sqrt{2}}(b|2\rangle-b^2|0\rangle),\quad
|\varphi_9\rangle=\frac{1}{\sqrt{2}}(b^2|2\rangle-b|0\rangle).
\end{aligned}
\end{equation*}

\end{widetext}

\end{document}